\documentclass[useAMS,usenatbib,psfig]{mn2e}
\usepackage{psfig}
\usepackage{times}

\title[The launching condition of a jet from an accretion disc] {The launching condition of a jet driven by the
magnetic field and radiation pressure of an accretion disc}
\author[X. Cao]
{Xinwu Cao\\Key Laboratory for Research in Galaxies and Cosmology,
Shanghai Astronomical Observatory, Chinese Academy of Sciences, \\80
Nandan Road, Shanghai, 200030, China; E-mail: cxw@shao.ac.cn }

\date{Accepted 2012 August 21. Received 2012 August 20; in original form 2012 June 26}

\pagerange{\pageref{firstpage}--\pageref{lastpage}} \pubyear{}

\begin{document}

\maketitle \label{firstpage}

\begin{abstract}
We find that the cold gas can be magnetically launched from the disc
surface with the help of the radiation pressure if the angular
velocity of the radiation pressure dominated accretion disc is
greater than a critical value, which decreases with increasing the
 disc thickness $H_{\rm d}/R$ (radiation pressure). This indicates
the force exerted by the radiation from the disc indeed helps
launching the outflow. The rotational velocity of the gas in the
 disc depends on the strength of the magnetic field threading the
 disc and the inclination $\kappa_0$ ($\kappa_0=B_z/B_r$) of the
field line at the disc surface. The launching condition for the cold
gas at the disc surface sets an upper limit on the magnetic field
strength, which is a function of the field line inclination
$\kappa_0$ and the disc thickness $H_{\rm d}/R$. This implies a more
strict constraint on the maximal jet power can be extracted from a
radiation pressure dominated accretion disc than that derived
conventionally on the equipartition assumption.
\end{abstract}


\section{Introduction}

The winds driven from the accretion disc through the magnetic field
lines threading the disc have been considered as promising
explanations for jets/outflows observed in different types of the
sources, such as, active galactic nuclei (AGNs), X-ray binaries, and
young stellar objects \citep*[see reviews
in][]{1996epbs.conf..249S,2000prpl.conf..759K,2007prpl.conf..277P,2010LNP...794..233S}.
In this model, the ordered magnetic field co-rotates with the gases
in the disc, and a small fraction of the gases are driven along the
field line by the centrifugal force \citep{1982MNRAS.199..883B}. The
jets are powered by the rotational kinetic energy of the disc
through the ordered field threading the accretion disc in this
scenario. A crucial ingredient of this model is the angle of the
field line inclined to the midplane of the disc being less than
$\sim 60^\circ$ is required for launching jets from the mid-plane of
the Keplerian cold disc \citep*{1982MNRAS.199..883B}. This critical
angle could be larger than $60^\circ$ for the accretion disc
surrounding a rapidly spinning black hole
\citep{1997MNRAS.291..145C}, which indicates that the spin of black
hole may help launching jets centrifugally by cold magnetized discs
\citep{1997MNRAS.291..145C,2010A&A...517A..18S}. Such analyses were
carried out based on two assumptions, the gas is cold (i.e., without
considering its internal energy), and the gas is initially launched
from the midplane of a Keplerian accretion disc.
Strictly speaking, the circular motion of the gases in the accretion
 disc always deviates from Keplerian motion in the presence of a
large-scale magnetic field, which usually exerts a radial force
against the gravitation of the central object, and the accretion
 disc threaded by ordered magnetic field lines is therefore always
sub-Keplerian
\citep{1998ApJ...499..329O,2001ApJ...553..158O,2002A&A...385..289C,2011ApJ...737...94C}.
The gases are dominantly accelerated by the centrifugal force, which
is proportional to $R\Omega(R_{\rm i})$ ($R_{\rm i}$ is the radius
of the magnetic field line footpoint, and $\Omega$ is the angular
velocity of the accretion disc). A strong magnetic field is helpful
for launching the gases, while the stronger field leads to a lower
circular velocity of the gas in the disc, which decreases the mass
loss rate in the outflow or even suppresses the outflow. Such
effects on the launch of the outflow have been extensively explored
in the previous works
\citep*[e.g.,][]{1998ApJ...499..329O,2001ApJ...553..158O,2002A&A...385..289C}.

It was also suggested that the outflow can be accelerated by the
radiation pressure of the disc
\citep*[e.g.,][]{1985ApJ...294...96S,1995ApJ...451..498M}.
\citet{2000ApJ...538..684P} performed numerical simulations on the
radiation-driven winds from a luminous Keplerian accretion disc
threaded by a strong large-scale ordered magnetic field. It is found
that the radiation force is essential in producing winds from the
disc if the thermal energy of the gas is low or the field lines make
an angle of greater than 60$^\circ$ with respect to the disc
midplane. The numerical simulations carried out in
\citet{2000ApJ...538..684P} are limited to the Keplerian accretion
discs. Recently, the global structure of the accretion discs and
outflows around black holes was investigated with
radiation-magnetohydrodynamic simulations
\citep*[e.g.,][]{2010PASJ...62L..43T,2011ApJ...736....2O}. Similar
investigations were carried out for the cases of massive young stars
by \citet{2011ApJ...742...56V}. It is found that the magnetic force,
together with the radiation force exerted by accretion discs, can
efficiently drive outflows from luminous accretion discs.

It is believed that the black holes are accreting at high rates in
narrow-line Seyfert I galaxies (NLS1s), some young radio galaxies,
and microquasars
\citep*[e.g.,][]{2000ApJ...536L...5S,2009ApJ...698..840C,2009ApJ...701L..95W,2004MNRAS.355.1105F}.
Most of NLS1s are radio-quiet, while a fraction of them are
radio-loud and some of them may possess relativistic jets
\citep*[e.g.,][]{2003ApJ...584..147Z,2006PASJ...58..777D,2008ApJ...685..801Y,2010AJ....139.2612G}.
For microquasars, for example, GRS 1915+105, relativistic jets are
present in their high-luminosity states \citep*[e.g.,
see][]{2004MNRAS.355.1105F}. The outflows/jets in these sources
accreting at high rates may probably be magnetically launched from
the radiation pressure-dominated accretion discs.

In this work, we explore the condition for cold gas driven by the
magnetic force and the radiation force from the surface of a
radiation pressure dominated accretion disc. The disc structure,
especially the circular motion velocity of the gases in the disc, is
altered in the presence of a strong large-scale ordered magnetic
field threading the disc, which is properly considered in this work.
We describe our analyses in Sections 2 and 3, and the results and
discussion are in Sections 4 and 5.

\section{Structure of an radiation pressure dominated accretion disc}

The accretion disc is assumed to be hydrodynamical equilibrium in
the vertical direction. For the gas pressure-dominated accretion
 disc, the gradient of the gas pressure is balanced with the gravity
and magnetic force due to the curvature of the field in the vertical
direction
\citep*[][]{1998ApJ...499..329O,2001ApJ...553..158O,2002A&A...385..289C,2011ApJ...737...94C}.
However, the situation is slightly different for the radiation
pressure-dominated accretion disc, the vertical component of
gravitational force of the black hole is balanced with the radiation
pressure of the disc and the vertical component of the magnetic
force at the disc surface $z=H_{\rm d}$ \citep{1989MNRAS.238..897L}.
The curvature of the field line at the disc surface is usually very
small, and the vertical component of the magnetic force can be
neglected in the estimate of the disc scale-height. The scale-height
of the disc can be estimated with
\begin{equation}
{\frac {GMH_{\rm d}}{(R_{\rm i}^2+H_{\rm d}^2)^{3/2}}}={\frac
{f_{\rm rad}\kappa_{\rm T}}{c}}, \label{vertical_1}
\end{equation}
which can be re-written as
\begin{equation}
{\frac {\tilde{H}_{\rm d}}{(1+\tilde{H}_{\rm d}^2)^{3/2}}}={\frac
{f_{\rm rad}\kappa_{\rm T}}{R_{\rm i}\Omega_{\rm K}^2 c}},
\label{vertical_2}
\end{equation}
where $f_{\rm rad}$ is the flux from the unit surface area of the
 disc, $\tilde{H}_{\rm d}=H_{\rm d}/R_{\rm i}$, and $\Omega_{\rm
K}=(GM/R_{\rm i}^3)^{1/2}$ is the Keplerian angular velocity at
$R_{\rm i}$.

In a geometrically thin accretion disc, the circular velocity is
almost Keplerian without magnetic fields. The circular motion of the
 disc becomes sub-Keplerian in the presence of magnetic fields, and
the angular velocity $\Omega$ of the disc can be calculated with
\begin{equation}
R_{\rm i}\Omega_{\rm K}^2-R_{\rm i}\Omega^2={\frac
{B_R^{S}B_{z}}{2\pi\Sigma_{\rm d}R_{\rm i}}}={\frac
{B_{z}^2}{2\pi\Sigma_{\rm d}R_{\rm i}\kappa_0}}, \label{Omega_1}
\end{equation}
where $\Sigma_{\rm d}$ is the surface density of the disc, $B_R^S$
and $B_z$ are the radial and vertical components of the field at the
disc surface $z=H_{\rm d}$ respectively, and $\kappa_0=B_z/B_R^S$.
{The radial gradient of radiation pressure in the disc is not
included in Equation (\ref{Omega_1}), which is negligible compared
with the magnetic force in the thin accretion disc if the magnetic
field is sufficient strong.}

The pressure of a radiation pressure dominated accretion disc at the
mid-plane is
\begin{equation}
p_{\rm d}={\frac {4\sigma}{3c}}T_{\rm c}^4,\label{p_d_1}
\end{equation}
where $T_{\rm c}$ is the central temperature of the disc. The flux
radiated from the unit area of the disc surface is given by
\begin{equation}
f_{\rm rad}={\frac {8\sigma T_{\rm c}^4}{3\Sigma_{\rm d}\kappa_{\rm
T}}}.\label{f_rad_1}
\end{equation}
Equation (\ref{p_d_1}) can therefore be re-written as
\begin{equation}
p_{\rm d}={\frac {\Sigma_{\rm d}\kappa_{\rm T}f_{\rm
rad}}{2c}}.\label{p_d_2}
\end{equation}
 Substituting Equations
(\ref{vertical_2}), (\ref{p_d_1}), and (\ref{f_rad_1}) into Equation
(\ref{Omega_1}), we have
\begin{equation}
1-\tilde{\Omega}^2={\frac {2\beta\tilde{H}_{\rm
d}}{\kappa_0(1+\tilde{H}_{\rm d}^2)^{3/2}}}, \label{Omega_2}
\end{equation}
where the dimensionless quantities $\tilde{\Omega}$, and $\beta$,
are defined as
\begin{equation}
\tilde{\Omega}={\frac {\Omega}{\Omega_{\rm K}}};~~~~~\beta={\frac
{B_z^2}{8\pi}}/p_{\rm d}.\label{quantities}
\end{equation}

\section{Launching condition for cold gas from the disc surface}


The condition for the cold gas can be launched from the midplane of
a Keplerian accretion disc was given in \citet{1982MNRAS.199..883B},
which is a good approximation for geometrically thin accretion discs
with weak magnetic fields, i.e., the rotation of the discs has not
been altered much by the field. The situation becomes complicated
for the gas driven from the surface of a real disc with finite
thickness in the presence of a strong magnetic field. In this case,
the rotation of the gas in the disc deviates significantly from the
Keplerian value (see discussion in Section 2).

The effective potential along the field line threading the accretion
disc with angular velocity $\Omega$ at radius $R_{\rm i}$ is
\begin{equation}
\Psi_{\rm eff}(R,~z)=-{\frac {GM}{(R^2+z^2)^{1/2}}}-{\frac
{1}{2}}\Omega(R_{\rm i})^2R^2, \label{potent_eff_0}
\end{equation}
without considering the radiation force exerted on the gas in the
outflow. It becomes
\begin{equation}
\Psi_{\rm eff}(R,~z)=-{\frac {GM}{(R^2+z^2)^{1/2}}}-{\frac
{1}{2}}\Omega(R_{\rm i})^2R^2-{\frac {f_{\rm rad}}{c}}\kappa_{\rm
T}z, \label{potent_eff_1}
\end{equation}
while the radiation pressure is considered, where $f_{\rm rad}$ is
the flux emitted from the unit surface area of the disc, and the
Thompson scattering cross-section $\kappa_{\rm T}=0.4~{\rm
g}^{-1}{\rm cm}^{-2}$. For the outflow from the inner region of the
disc, the gases are almost completely ionized and the Thompson
scattering cross-section is therefore a good approximation. The
analysis carried out in this work can be applied to the outflow
driven by the radiation pressure due to the line absorption if the
Thompson scattering cross-section is replaced by the line opacity,
{though line driven outflows are usually from the outer region of
the disc}
\citep*[e.g.,][]{1985ApJ...294...96S,1995ApJ...451..498M,2000ApJ...543..686P}.

Substituting Equation (\ref{vertical_2}) into Equation
(\ref{potent_eff_1}), the effective potential along the field line
threading the accretion disc with angular velocity $\Omega$ at
radius $R_{\rm i}$ can be re-written in dimensionless form,
\begin{displaymath}
\tilde{\Psi}_{\rm eff}(r,~\tilde{z})={\frac {\Psi_{\rm
eff}(R,~z)}{R_{\rm i}^2\Omega_{\rm K}^2}}
\end{displaymath}
\begin{equation}
~~~~~~~~~~~~~=-{\frac {1}{(r^2+\tilde{z}^2)^{1/2}}}-{\frac
{1}{2}}\tilde{\Omega}^2r^2-{\frac {\tilde{z}\tilde{H}_{\rm
d}}{(1+\tilde{H}_{\rm d}^2)^{3/2}}}, \label{potent_eff_2}
\end{equation}
where $r=R/R_{\rm i}$, and $\tilde{z}=z/R_{\rm i}$.
Differentiating Equation
(\ref{potent_eff_2}), we have
\begin{equation}
{\frac {d\tilde{\Psi}_{\rm eff}(r,~\tilde{z})}{dr}}={\frac
{r+\tilde{z}\kappa_0}{(r^2+\tilde{z}^2)^{3/2}}}-r\tilde{\Omega}^2-{\frac
{\tilde{H}_{\rm d}\kappa_0}{(1+\tilde{H}_{\rm d}^2)^{3/2}}},
\label{dpsi_eff_dr1}
\end{equation}
which reduces to
\begin{equation}
{\frac {d\tilde{\Psi}_{\rm eff}(r,~\tilde{z})}{dr}}\left
\arrowvert_{r=1,\tilde{z}=\tilde{H}_{\rm d}}={\frac
{1}{(1+\tilde{H}_{\rm d}^2)^{3/2}}}-\tilde{\Omega}^2\right.,
\label{dpsi_eff_dr2}
\end{equation}
at the disc surface, $r=1$ and $\tilde{z}=\tilde{H}_{\rm d}$. This
means that the cold gas can be magnetically launched from the disc
surface with the help of radiation pressure only if the condition,
\begin{equation}
{\frac {d\tilde{\Psi}_{\rm eff}(r,~\tilde{z})}{dr}}\left
\arrowvert_{r=1,\tilde{z}=\tilde{H}_{\rm d}}\le 0\right.,
\label{condi_coldgas1}
\end{equation}
is satisfied, which requires
\begin{equation}
\tilde{\Omega}\ge {\frac {1}{(1+\tilde{H}_{\rm d}^2)^{3/4}}}.
\label{condi_coldgas2}
\end{equation}
Substituting Equation (\ref{Omega_2}) into Equation
(\ref{condi_coldgas2}), the condition becomes
\begin{equation}
\beta\le{\frac {\kappa_0}{2\tilde{H}_{\rm d}}}[(1+\tilde{H}_{\rm
d}^2)^{3/2}-1],\label{condi_coldgas3}
\end{equation}
for the cold gas being able to leave the disc surface along the
field line. {In order to clarify the role of the radiation force in
launching the jets/outflows, we analyze two special cases: A $f_{\rm
rad}\rightarrow 0$ and $\tilde{\Omega}\rightarrow 1$; and, B $f_{\rm
rad}$ is not considered and $\tilde{\Omega}<1$. }

\subsection{Case A: $f_{\rm rad}\rightarrow 0$ and $\Omega/\Omega_{\rm K}\rightarrow
1$}

In our model, the circular velocity of the accretion discs deviates
from the Keplerian value in the presence of magnetic fields. A lower
$\beta$ leads to a larger angular velocity $\tilde{\Omega}$ for a
fixed value of $\kappa_0$ (see Equation \ref{Omega_2}), which helps
launching the outflow. The results derived here are different from
that derived in \citet{1982MNRAS.199..883B} for the gas driven from
the midplane of a Keplerian disc. {One may expect that our
calculation can reproduce the Blandford-Payne's result on the
critical angle of the field line with respect to the disc plane,
below which the cold gas can be launched from the mid-plane of a
Keplerian disc. For a radiation pressure dominated accretion disc,
the gravity of the black hole is balanced with the radiation
pressure of the disc in the vertical direction, which means that the
disc thickness $H_{\rm d}\rightarrow 0$ in the limit of $f_{\rm
rad}\rightarrow 0$ (see Equation \ref{vertical_2}), and
$\tilde{\Omega}\rightarrow 1$ is automatically satisfied (see
Equation \ref{Omega_2}). Thus, our model calculation corresponds to
the case considered in \citet{1982MNRAS.199..883B} for the cold gas
from the midplane of a Keplerian disc in the limit of
$\tilde{H}_{\rm d}\rightarrow 0$. We find that ${d\tilde{\Psi}_{\rm
eff}(r,~\tilde{z})}/{dr}=0$ if $\tilde{H}_{\rm d}=0$ (see Equation
\ref{dpsi_eff_dr2}). It implies that the gas can always be in
equilibrium independent of the field line inclination $\kappa_0$.
This is the same as that in \citet{1982MNRAS.199..883B}. Therefore,
one has to calculate the second derivative of the effective
potential to check the stability of equilibrium,
\begin{equation}
{\frac {d^2\tilde{\Psi}_{\rm eff}(r,~\tilde{z})}{dr^2}}={\frac
{-2r^2-6r\tilde{z}\kappa_0-2\tilde{z}^2\kappa_0^2+\tilde{z}^2+r^2\kappa_0^2}{(r^2+\tilde{z}^2)^{5/2}}}
-\tilde{\Omega}^2,\label{d2psi_eff_dr21}
\end{equation}
which is required to be negative at $r=1$ and
$\tilde{z}=\tilde{H}_{\rm d}=0$ for the cold gas can be launched
from the midplane of the disc. This leads to
\begin{equation}
\kappa_0<\kappa_{0,\rm crit}=\sqrt{3},\label{kappa_crit_bp1}
\end{equation}
which reaches the same result given in \citet{1982MNRAS.199..883B}.
Our calculations show that gas at the disc surface is not in
equilibrium state when the angular velocity is greater than a
critical value for the disc with finite thickness, and therefore the
launch condition for the cold gas at the disc surface can be derived
by the first derivative of the effective potential
${d\tilde{\Psi}_{\rm eff}(r,~\tilde{z})}/{dr}<0$ at $z=H_{\rm d}$. }

\subsection{Case B: $f_{\rm rad}$ is not considered and
$\Omega<\Omega_{\rm K}$}

{In most of the previous work, the effect of the radiation force
exerted by the accretion disc on the outflow/jet has not been
considered, while this effect is included in this work. We leave out
the radiation pressure term,} and re-analyze the effective potential
along the field line. {The analysis carried out here is in principle
inconsistent with the assumption of the radiation pressure dominated
accretion discs. However, the analysis is only limited to the
condition of the gas being able to leave the system, which is almost
independent of the disc structure. It illustrates the role of the
radiation pressure in launching the outflow from the surface of the
disc.} The effective potential along the field line threading the
accretion disc with angular velocity $\Omega$ at radius $R_{\rm i}$
can be written in dimensionless form,
\begin{equation}
\tilde{\Psi}_{\rm eff}(r,~\tilde{z})={\frac {\Psi_{\rm
eff}(R,~z)}{R_{\rm i}^2\Omega_{\rm K}^2}}=-{\frac
{1}{(r^2+\tilde{z}^2)^{1/2}}}-{\frac {1}{2}}\tilde{\Omega}^2r^2,
\label{potent_eff_2nr}
\end{equation}
where $r=R/R_{\rm i}$, $\tilde{z}=z/R_{\rm i}$, and, {the term due
to the radiation force in Equation (\ref{potent_eff_2}) is omitted}.
Differentiating Equation (\ref{potent_eff_2nr}), the condition for
the cold gas being able to leave the disc surface is available,
\begin{equation}
{\frac {d\tilde{\Psi}_{\rm eff}(r,~\tilde{z})}{dr}}\left
\arrowvert_{r=1,\tilde{z}=\tilde{H}_{\rm d}}={\frac
{r+\tilde{H}_{\rm d}\kappa_0}{(1+\tilde{H}_{\rm
d}^2)^{3/2}}}-\tilde{\Omega}^2\le 0\right., \label{condi_coldgasnr}
\end{equation}
{and we derive the lower limit on the angular velocity of the disc
as}
\begin{equation}
\tilde{\Omega}\ge {\frac {(1+\tilde{H}_{\rm
d}\kappa_0)^{1/2}}{(1+\tilde{H}_{\rm d}^2)^{3/4}}}.
\label{condi_coldgasnr2}
\end{equation}
{The constraint of the magnetic field strength is available,}
\begin{equation}
\beta\le{\frac {\kappa_0}{2\tilde{H}_{\rm d}}}[(1+\tilde{H}_{\rm
d}^2)^{3/2}-1-\tilde{H}_{\rm d}\kappa_0], \label{condi_coldgasnr3}
\end{equation}
{by substituting Equation (\ref{Omega_2}) into Equation
(\ref{condi_coldgasnr2}). It is found that the critical angular
velocity of the disc, above which the cold gas can be launched from
the disc surface, should be greater than the Keplerian value when
$\kappa_0$ is sufficiently large (see Equation
\ref{condi_coldgasnr}), if the role of the radiation force on the
outflow is not considered. In the case of the radiation force being
properly considered, we find that the critical angular velocity of
the disc is always lower than the Keplerian value (see Equation
\ref{condi_coldgas2}). It implies that the radiation force can help
for launching the gas from the disc surface. We note that
${d\tilde{\Psi}_{\rm eff}(r,~\tilde{z})}/{dr}=0$ when
$\tilde{H}_{\rm d}=0$ from Equation (\ref{condi_coldgasnr}). As
discussed above, the second derivative of the effective potential
$d^2\tilde{\Psi}_{\rm eff}/{dr^2}<0$ at $r=1$ and $\tilde{H}_{\rm
d}=0$ is required for the cold gas being able to leave. The last
term in the first derivative of the effective potential (Equation
\ref{dpsi_eff_dr1}) is left out when the role of the radiation force
on the outflow is not considered. This term remains constant along
the field line, and the second derivative of the effective potential
has the same form as Equation (\ref{d2psi_eff_dr21}). Thus, our
analysis without considering radiation pressure can also reproduce
the same result in \citet{1982MNRAS.199..883B} when $\tilde{H}_{\rm
d}\rightarrow 0$. }

\section{Results}

\begin{figure}
\centerline{\psfig{figure=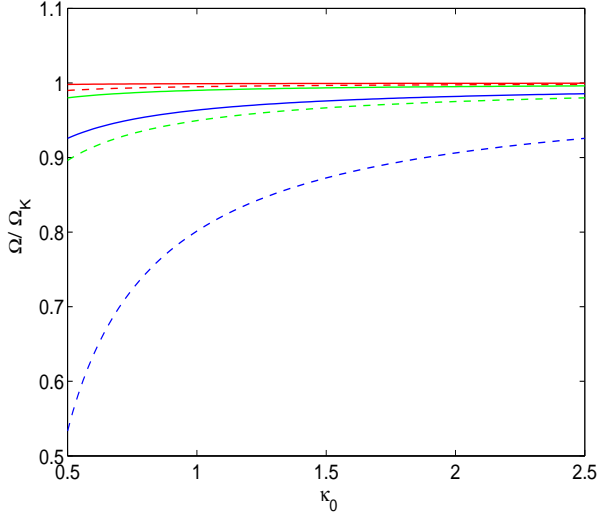,width=8.0cm,height=7.0
cm}} \caption{The angular velocity of the accretion disc as
functions of $\beta$ and $\kappa_0$. The solid lines represent the
results with $\beta=0.1$, while the dashed lines are for
$\beta=0.5$. The color lines correspond the results with different
values of the disc scale height: $\tilde{H}_{\rm d}=0.01$ (red),
$0.1$ (green), and $0.5$ (blue). } \label{kappa_omega}
\end{figure}

As discussed in Section 2, the angular velocity of the gas in the
accretion disc is dependent on the magnetic field strength $\beta$
and the inclination of the field line $\kappa_0$ at the disc
surface. We plot the angular velocities of the accretion disc as
functions of $\beta$ and $\kappa_0$ with different values of
$\tilde{H}_{\rm d}$ in Figure \ref{kappa_omega}. {It is found that
the angular velocity $\tilde{\Omega}$ increases with field line
inclination $\kappa_0$ at the disc surface. For the fixed value of
$\kappa_0$, the angular velocity $\tilde{\Omega}$ increases with
magnetic field strength $\beta$. }

\begin{figure}
\centerline{\psfig{figure=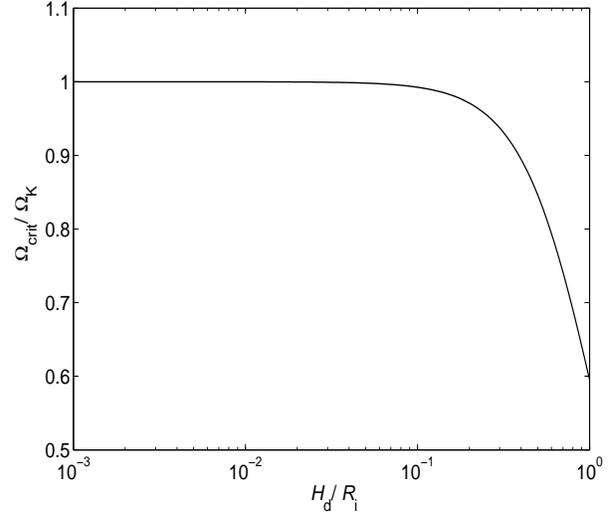,width=8.0cm,height=7.0 cm}}
\caption{The condition of the cold gas can be launched from the disc
surface (see Equation \ref{condi_coldgas2}).} \label{hd_omega}
\end{figure}

\begin{figure}
\centerline{\psfig{figure=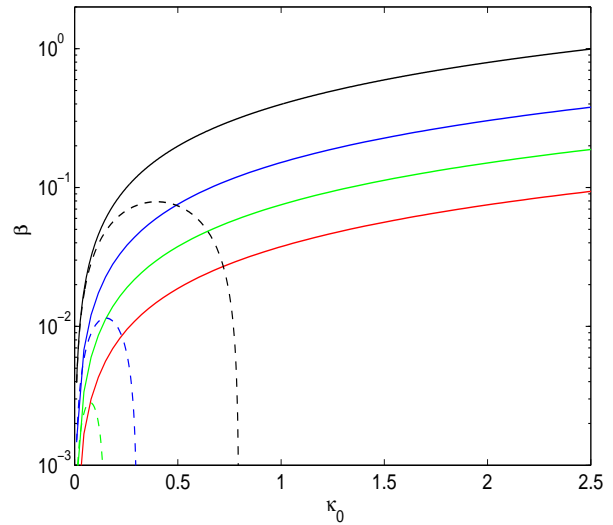,width=8.0cm,height=7.0
cm}} \caption{The solid lines represent the conditions of the cold
gas can be launched from the disc surface as functions of $\kappa_0$
and $\beta$(see Equation \ref{condi_coldgas3}), while the dashed
lines are the results calculated without considering the effect of
radiation pressure on the launch of the outflow/jet (see the text in
Section 2.2). The colored lines are for the different disc
thickness, $\tilde{H}_{\rm d}=0.05$(red), $0.1$(green), $0.2$(blue),
and $0.5$(black). } \label{kappa_beta_crit}
\end{figure}

The launch of the cold gas in the disc is governed by the effective
potential barrier, and the cold gas can be launched from the disc
surface along the field line if the angular velocity of the gas in
the disc is larger than a critical value (see Equation
\ref{condi_coldgas2}), which is plotted in Figure \ref{hd_omega}.
{The critical angular velocity $\tilde{\Omega}_{\rm crit}$ decreases
with increasing disc thickness $\tilde{H}_{\rm d}$. } As
$\tilde{\Omega}$ is a function of $\beta$, $\kappa_0$ and
$\tilde{H}_{\rm d}$, the conditions of the cold gas can be launched
from the disc surface are available as functions of $\kappa_0$ and
$\beta$ with given $\tilde{H}_{\rm d}$ (see Figure
\ref{kappa_beta_crit}). {There are upper limits on the magnetic
field strength, which increase with the field inclination $\kappa_0$
and disc thickness $\tilde{H}_{\rm d}$.} For comparison, we also
plot the results without considering the effect of radiation
pressure from the disc in the same figure. {We find more strict
constraints on the magnetic field strength. The cold gas can be
launched from the disc surface only if the field line is bent close
to the disc surface.   }


\section{Discussion}

We investigate the launch of the gas from the disc surface by the
field lines threading a radiation pressure-dominated accretion
 disc, in which the effect of the radiation pressure from the
accretion disc is considered. The launch of the outflow is sensitive
to the rotational angular velocity of the gas in the disc, which is
determined by the balance between the gravity of the black hole and
the magnetic force. The angular velocity of the disc is a function
of the field line inclination $\kappa_0$ at the disc surface and
$\beta$ (the ratio of the magnetic pressure to the radiation
pressure in the disc), if the relative disc thickness
$\tilde{H}_{\rm d}$ is specified (see Figure \ref{kappa_omega}). It
is not surprising that the angular velocity of the gas in the disc
$\tilde{\Omega}$ decreases with increasing field strength $\beta$,
if the values of all other parameters are fixed. The angular
velocity $\tilde{\Omega}$ becomes smaller for the field line
inclined at a smaller angle with respect to the disc surface, which
is caused by a stronger magnetic force for a smaller $\kappa_0$. We
find that the angular velocity $\tilde{\Omega}$ decreases with
increasing relative disc thickness $\tilde{H}_{\rm d}$ for the same
values of $\beta$ and $\kappa_0$.

Similar to the discussion in \citet{1982MNRAS.199..883B}, we
investigate the condition for the cold gas can be launched from the
 disc surface. In the presence of magnetic fields, the rotational
velocity of the gas in the disc is sub-Keplerian, which makes the
gas being more difficult to leave the disc surface than the
Keplerian case \citep*[][]{1998ApJ...499..329O,2001ApJ...553..158O}.
In this work, we include the effect of radiation pressure in
launching the outflow. We find that the cold gas can be magnetically
driven from the disc surface if the angular velocity of the disc
greater than a critical value $\tilde{\Omega}_{\rm crit}$ (see
Figure \ref{hd_omega}), which is a function of the disc thickness
$\tilde{H}_{\rm d}$. For the radiation pressure dominated accretion
 disc, the disc thickness $\tilde{H}_{\rm d}$ describes the
importance of radiation pressure (see Equation \ref{vertical_2}). It
is found that the critical value of $\tilde{\Omega}_{\rm crit}$
decreases with increasing $\tilde{H}_{\rm d}$, which indicates that
the force exerted by the radiation from the disc indeed helps
launching the outflow. The angular velocity of the gas in the disc
is determined by the field strength and the inclination of the field
line at the disc surface, and therefore the condition of the cold
gas can be launched becomes a function of $\kappa_0$ and $\beta$
(see Figure \ref{kappa_beta_crit}). For the given magnetic field
line inclination $\kappa_0$, the cold gas can be driven from the
disc surface only if the field strength is lower than a certain
value (see Equation \ref{condi_coldgas3}). This is because a higher
$\beta$ leads to a lower angular velocity $\tilde{\Omega}$ for a
fixed value of $\kappa_0$. It is found that the launch of the cold
outflow is suppressed if the magnetic field is too strong, which is
due to the decrease of $\tilde{\Omega}$ with increasing $\beta$. For
a relatively thick disc, the upper limit on the field strength could
be high because the cold flow can be launched from the disc surface
with a relatively low $\tilde{\Omega}$ with the help of the
radiation force. This implies a more strict constraint on the
maximal jet power can be extracted from a radiation pressure
dominated accretion disc than that derived conventionally on the
equipartition assumption.

In Figure \ref{kappa_beta_crit}, we also plot the results without
considering the effect of the radiation pressure exerted by the
accretion disc, i.e., the radiation pressure term is left out in the
effective potential, in order to explore the role of the radiation
pressure in launching the ouflow. It is found that the cold gas can
be launched from the disc surface only if $\kappa_0$ is small, i.e.,
the angle of the field line inclined with respect to the disc
surface is small, if the radiation pressure term is not included.
This implies that the role of the radiation pressure in launching an
outflow is important even for thin accretion discs. The analyses in
this work are carried out on the assumption of a radiation pressure
dominated accretion disc, which is justified in the inner region of
the accretion disc if the accretion rate is not very low
\citep*[see, e.g.,][]{1989MNRAS.238..897L}. The calculations are
done only for the cold gas driven from the disc surface, because the
outflow can be efficiently driven from the place near the disc
surface where the magnetic pressure is dominant over the gas
pressure. The structure of the disc in the presence of a magnetic
field is complicated, which should be considered by solving the
differential equations governing the vertical structure of the disc
and its magnetic field consistently. Such calculations without
considering the effect of radiation force have already been given in
the previous works \citep*{1998ApJ...499..329O,2001ApJ...553..158O}.
The similar investigations incorporated with the effect of radiation
force are beyond the scope of this paper, which will be reported in
our future work.

{In most of the MHD simulations, the accretion flows have relative
large thickness, and the gas is launched from the magnetic pressure
dominated region near the disc surface
\citep*[e.g.,][]{1999ApJ...522..727K,2003ApJ...592.1060D,2009MNRAS.397.2087M,2012MNRAS.423.3083M}.
Their the rotational velocities deviate from the Keplerian value
mainly due to the magnetic force and gas pressure gradient in the
radial direction
\citep*[e.g.,][]{2000ApJ...528..462H,2003ApJ...592..767P}. Our
present analysis on the gas launched from the disc surface describes
the situation more akin to the numerical simulations than that in
\citet{1982MNRAS.199..883B}, which is valid only for the gas
launched from the midplane of a Keplerian disc. The radiation of the
accretion flows are calculated in some previous works based on the
structure given by MHD simulations, which are used to explain the
observations
\citep*[e.g.,][]{2007A&A...474....1M,2009ApJ...706..497M}. However,
the radiation from the accretion flows has not been considered in
these MHD simulations. The role of the radiation pressure on the
magnetically outflows has been explored in a few MHD simulations
\citep{2000ApJ...538..684P,2003ApJ...585..406P}, but limited to the
Keplerian disc case. The recent radiation-magnetohydrodynamic
simulations on the accretion discs and outflows around black holes
show that the radiation force helps launch outflows from luminous
accretion discs \citep*{2010PASJ...62L..43T,2011ApJ...736....2O},
which is qualitatively consistent with our analytic results.}

\section*{Acknowledgments}
I thank the referee for his/her helpful comments, and Qingwen Wu for
helpful discussion. This work is supported by the National Basic
Research Program of China (grant 2009CB824800), the NSFC (grants
11173043, 11121062 and 10833002), and the CAS/SAFEA International
Partnership Program for Creative Research Teams (KJCX2-YW-T23).

\end{document}